\documentclass{raa}
\usepackage{amsmath}

\usepackage{graphicx,times}             
\input{epsf.sty}                        

\begin{document}
 \title{Exploring the sources of p-mode frequency shifts in the CoRoT target HD~49933}

   \volnopage{Vol.0 (200x) No.0, 000--000}      
   \setcounter{page}{1}          

   \author{Zhie Liu
      \inst{1}
   \and Shaolan Bi
      \inst{1}
   \and Wuming Yang
      \inst{1}
    \and Tanda Li
    \inst{2}
   \and Kang Liu
       \inst{1}
   \and Zhijia Tian
      \inst{1}
   \and Zhishuai Ge
      \inst{1}
      \and Jie Yu
      \inst{1}
   }

   \institute{Deparment of Astronomy, Beijing Normal University,
             Beijing 100875, China; {\it zhieliu@mail.bnu.edu.cn}\\
             \and
            Key Laboratory of Solar Activity, National Astronomical Observatories, Chinese Academy of Science,
             Beijing 100012, China;\\
   }

   \date{Received~~2009 month day; accepted~~2009~~month day}

\abstract{
The oscillations of the solar-like star HD~49933 have been observed thoroughly by CoRot. Two dozens of frequency shifts, which are closely related with the change in magnetic activity, have been measured. To explore the effects of the magnetic activity on the frequency shifts, we calculate frequency shifts for the radial and $l = 1$ p-modes of HD~49933 with the general variational method, which evaluates the shifts using a spatial integral of the product of a kernel and some sources. The theoretical frequency shifts well reproduce the observation. The magnitudes and positions of the sources are determined according to the  $\chi^2$ criterion. We predict the source that contributes to both $l = 0$ and $l = 1$ modes is located at $0.48 - 0.62$Mm below the stellar surface. In addition, based on the assumption that $A_{0}$ is proportional to the change in the MgII activity index $\Delta{i}_{MgII}$, we obtained that the change of MgII index between minimum and maximum of HD~49933 cycle period is about 0.665. The magnitude of the frequency shifts compared to the Sun already told us that
HD~49933 is much more active than the Sun, which is further confirmed in this paper. Furthermore, our calculation on the frequency shifts of $l = 1$ modes indicates the variation of turbulent velocity in the stellar convective zone may be an important source for the $l = 1$ shifts.
 \keywords{stars:
individual: HD~49933 --- stars: evolution
--- stars: oscillation --- stars: modelling}}

   \authorrunning{Zhie Liu, Shaolan Bi \& Wuming Yang, et al. }            
   \titlerunning{Modeling the p-mode frequency shifts in HD~49933}  

   \maketitle

\section{Introduction}
\label{sect:intro}

HD~49933, also known as HR 2530 and HIP 32851, is an F5V main-sequence star with a surface rotation period of 3.5 days. It has a temperature ranging between 6450 $\pm$ 75K (Kallinger et al. \cite{Kallinger10}) and 6780 $\pm$ 130K (Bruntt et al. \cite{Bruntt04}), a $\log(L/L_{\odot})$ between 4.24 $\pm$ 0.13 (Bruntt et al. \cite{Bruntt08}) and 4.3 $\pm$ 0.2 (Bruntt et al. \cite{Bruntt04}), and a radius of 1.42 $\pm$ 0.04 $R(R_{\odot})$. Its metallicity, ranging from -0.46 $\pm$ 0.08 (Bruntt et al. \cite{Bruntt08}) to -0.30 $\pm$ 0.11 (Bruntt et al. \cite{Bruntt04}), is slightly metal poor compared to the Sun and to Procyon. All these characteristics are summarized in table \ref{Tab:sobs}. Its oscillation, with a first observation in Doppler velocity by Mosser et al
(\cite{Mosser05}), was observed three times by CoRoT in recent years but the last run is still under processing by the CoRoT team. The first observed time span was 60-day at the beginning of 2007 during the first CoRoT run (IRa01) while the second was 137-day in 2008 during the first CoRoT long duration run (LRa01). Garc\'{\i}a et al. (\cite{Garcia10}) analyzed such two sets of data and discovered that the p-mode frequencies and amplitudes of HD~49933 varied with the magnetic activity with a period of at least 120 days. Subsequently, Salabert et al. (\cite{Salabert11}) analyzed the second set of data by dividing the 137-day light curve into two subseries corresponding to periods of low- and high-stellar activity based on the work of Garc\'{\i}a et al. (\cite{Garcia10}) and extracted 24 frequencies and their shifts for $l = 0$ and $l = 1$ modes using a local maximum-likelihood fitting analysis, 12 for $l = 0$ modes, and 12 for $l = 1$ modes, which are listed in table \ref{Tab:shifts} (Salabert et al. Private communication).

We can find from table \ref{Tab:shifts} that, for $l = 0$ modes, most frequency shifts are changed in the range of 1-3 $\mu$Hz, quite high compared to the frequency shifts of the Sun and $\beta$Hyi. In addition, the frequency shifts of HD~49933 reach a maximal value of about 3 $\mu$Hz around 2100 $\mu$Hz. For the frequencies larger than 2100 $\mu$Hz, the variation of the p-mode frequency shifts indicates a downturn followed by an upturn for both $l = 0$ and $l = 1$ modes. Such a frequency dependence of the frequency shifts measured in HD~49933 is comparable with the one observed in the Sun (Salabert et al., \cite{Salabert04}), suggesting the solar-like star HD~49933 could have a similar physical mechanism driving the frequency shifts as the ones taking place in the Sun, which is thought to arise from changes in the outer layers due to its magnetic activity (Salabert et al. \cite{Salabert11}).

\begin{table}
\begin{center}
\renewcommand\arraystretch{1.0}
\caption[]{ Observational and theoretical data for HD~49933.}\label{Tab:sobs}
 \begin{tabular}{lllll}
  \hline\hline\noalign{\smallskip}
  T$_{eff}$(K) & $\log($L/L$_{\odot})$ & $[Fe/H]$ &R(R$_{\odot})$& Ref.\\
  \hline\noalign{\smallskip}
                  6570 $\pm$ 60   &   & -0.44 $\pm$ 0.03 &  & VWA\\
                  6450 $\pm$ 75 &   & &  &Kallinger et al.(\cite{Kallinger10}) \\
                  6780 $\pm$ 130 & 4.24 $\pm$ 0.13 &-0.46  $\pm$ 0.08 &  & Bruntt et al.(\cite{Bruntt08})\\
                  6735 $\pm$ 53 & 4.26 $\pm$ 0.08  & -0.37 $\pm$ 0.03$^{\ast}$ &  & Gillon \& Magain(\cite{Gillon06})\\
                  6780 $\pm$ 70 & 4.3 $\pm$ 0.2    & -0.30 $\pm$ 0.11 &  & Bruntt et al.(\cite{Bruntt04})\\
                       &  & &1.42 $\pm$ 0.04$^{\ast}$  &Bigot et al.(\cite{Bigot11})\\
  \hline\noalign{\smallskip}
  \hline\noalign{\smallskip}
\end{tabular}
\end{center}
\tablenotes{\ast}{0.86\textwidth}{Liu et al. (\cite{Liu13}) used data}
\end{table}

\begin{table}
\begin{center}
\renewcommand\arraystretch{1.0}
\caption[]{ Observed frequencies and their shifts of the star HD49933 (Salabert et al. \cite{Salabert11})$^\ast$.}
 \label{Tab:shifts}
\begin{tabular}{p{2.00cm}cccc}
  \hline\hline\noalign{\smallskip}
      Frequency    & Frequency shift & Frequency    & Frequency shift \\
      ($\mu$Hz)   &     ($\mu$Hz)        & ($\mu$Hz)& ($\mu$Hz) \\
  \hline\noalign{\smallskip}
   l = 0                          &             &    l = 1  &  & \\
  \hline\noalign{\smallskip}
              1544.69 $\pm$ 0.83 & 0.753 $\pm$ 1.007 & 1500.54 $\pm$ 0.94&-0.336 $\pm$ 0.848\\
              1631.10 $\pm$ 0.22 & 0.780 $\pm$ 1.123 &  1586.62 $\pm$ 0.61&-0.111 $\pm$ 0.666\\
              1714.49 $\pm$ 0.61 & -0.606 $\pm$ 2.108& 1670.48 $\pm$ 0.81& 1.139 $\pm$ 0.948\\
              1799.75 $\pm$ 1.03 & 2.073 $\pm$ 1.508 & 1755.30 $\pm$ 0.78&-0.496 $\pm$ 0.981\\
              1884.82 $\pm$ 0.59 & 0.815 $\pm$ 2.050 & 1840.68 $\pm$ 0.79& 1.657 $\pm$ 0.853\\
              1972.73 $\pm$ 1.14 & 1.344 $\pm$ 1.158 & 1928.13 $\pm$ 1.48& 0.233 $\pm$ 0.750\\
              2057.82 $\pm$ 0.96 & 3.059 $\pm$ 1.674 & 2014.38 $\pm$ 0.93& 0.717 $\pm$ 0.992\\
              2147.10 $\pm$ 1.05 & 2.808 $\pm$ 0.863 & 2101.58 $\pm$ 1.67& 1.091 $\pm$ 1.021\\
              2236.46 $\pm$ 0.39 & 1.868 $\pm$ 1.658 & 2190.81 $\pm$ 2.32&-0.578 $\pm$ 1.095\\
              2322.10 $\pm$ 1.66 & 1.430 $\pm$ 2.193 & 2277.89 $\pm$ 1.29&-3.338 $\pm$ 1.826\\
              2408.56 $\pm$ 0.83 & 2.174 $\pm$ 0.989 & 1500.54 $\pm$ 0.94&-0.336 $\pm$ 0.848\\
              2495.76 $\pm$ 3.34 & 11.444 $\pm$ 2.812 & 1586.62 $\pm$ 0.61&-0.111 $\pm$ 0.666\\
                 \noalign{\smallskip}\hline\hline
\end{tabular}
\end{center}
\tablenotes{\ast}{0.86\textwidth}{Data gently provided through private communication.}
\end{table}

\begin{table}
\begin{center}
\renewcommand\arraystretch{1.0}
\caption[]{ Parameters of the Sun and $\beta$Hyi.}
 \label{Tab:sunb}
\begin{tabular}{ccccccc}
  \hline\noalign{\smallskip}
  \hline\noalign{\smallskip}
      star     & $A_{0}$& Ref. & $\Delta i_{MgII}$ &Ref. & Frequency shift($\mu$Hz) &Ref.\\
      \hline
      Sun      & 0.3116& (1)   &   0.0135  &(1)  & most $<$ 0.8 &(1) \\
    $\beta$Hyi & 0.33  & (1)   &   0.015   &(1)  & 0.1 $\pm$ 0.4&(2) \\
     HD~49933  & 14.63$^{\ast}$ &       &   0.665$^{\ast}$    &     & most 1-3     &(3) \\
                 \noalign{\smallskip}\hline\hline
\end{tabular}
\end{center}
\tablerefs{0.86\textwidth}{
(1) Matcalfe et al. \cite{Metcalfe07};
(2) Bedding et al. \cite{Bedding07};
(3) Salabert et al. \cite{Salabert11}.
 }
 \tablenotes{\ast}{0.86\textwidth}{Results in this paper.}
\end{table}

In dynamo modeling, frequency shifts are thought to arise from either changes in the near-surface propagation speed due to a direct magnetic perturbation (Goldreich et al. \cite{Goldreich91}), or a slight decrease in the radial component of the turbulent velocity in the outer layers and the associated temperature changes (Dziembowski \& Goode \cite{Dziembowski04}, \cite{Dziembowski05}). Metcalfe et al. (\cite{Metcalfe07}) developed a method for predicting frequency shifts of solar-like stars based on scaling the measured p-mode frequency variations and changes of the chromospheric activity indices in the Sun. To forecast the frequency shifts,  Metcalfe et al. (\cite{Metcalfe07}) resorted to MgII index of the star and assumed that the relationship between variation of MgII index and the source strength in the Sun is also valid in other stars. Specific forecast was made for the radial modes in the subgiant $\beta$ Hyi, and their calculation results were consistent with the observed shifts. This work was generalized to nonradial modes by Dziembowski (\cite{Dziembowski07}), with an additional assumption that the Sun and other stars share the same butterfly diagram pattern.

In our previous work (Liu et al. \cite{Liu13}) we used the small frequency separation ratios $r_{01}$ and $r_{10}$ to constrain the evolution parameters of the stellar models and determine the size of the convective core and the extent of overshooting for HD~49933. In the present work, we will utilize the stellar models we have obtained in Liu et al. (\cite{Liu13}), with the method developed by Metcalfe et al. (\cite{Metcalfe07}) to study the observed frequency shifts of HD~49933.

In Sect. 2 we outline the general variational method for modeling the frequency shifts. Then we apply this method to study the shifts of radial and nonradial modes of HD~49933 in Sect. 3. Finally, we discuss our results and give the conclusions in Sect. 4.

\section{Method}
\label{sect:Method}
\subsection{General foumulations}
In order to evaluate the activity-related frequency shifts, we use a general variational expression giving by Metcalfe et al. (\cite{Metcalfe07}),

\begin{equation}
\Delta\nu_{nlm} = \frac{\int{d^3\vec{r}\mathcal{K}_{nlm}\mathcal{S}}}{2I_{nl}\nu_{nlm}}
\label{eq:01}
\end{equation}

where
\begin{equation}
I_{nl} = \int^R_{0}\rho[\xi^2_{r}+\Lambda\xi^2_{h}]r^2dr = R^5\overline{\rho}\widetilde{I}_{nl}
\label{eq:02}
\end{equation}
is the mode inertia. $\Lambda = l(l+1)$ and $\overline{\rho}$ is the stellar mean density. The dimensionless mode inertia, $\widetilde{I}_{nl}$, is defined as
\begin{equation}
\widetilde{I}_{nl} = \int^1_{0}\widetilde{\rho}[y^2+\Lambda{z^2}]x^4dx
\label{eq:03}
\end{equation}
where $x = \frac{r}{R}$, $\widetilde{\rho} = \frac{\rho}{\overline{\rho}}$, $y = \frac{\xi_{r}}{r}$, $z = \frac{\xi_{h}}{r}$ are corresponding dimensionless quantities.
Since all the derived kernels have leading terms proportional to $|div\vec{\xi}_{nlm}(\vec{r})|^2$, for simplicity Metcalfe et al. (\cite{Metcalfe07}) adopted the common kernel
\begin{equation}
\mathcal{K}_{nlm} = |div\vec{\xi}_{nlm}(\vec{r})|^2 = q_{nl}(D)|Y^{m}_{l}|^2
\label{eq:ker}
\end{equation}
where $D$ denotes the depth below the photosphere, and $Y^{m}_{l}$ is the spherical harmonics. It is easily verified that
\begin{equation}
q_{j}(D) = [\frac{1}{r^2}\frac{\partial}{\partial{r}}(r^2\xi_{r}(r))-\frac{1}{r}\xi_{h}(r)\Lambda]^2.
\label{eq:05}
\end{equation}
The following simple form of source was assumed in Metcalfe et al. (\cite{Metcalfe07}):
\begin{equation}
\mathcal{S}_{k}(D) = 1.5\times10^{-11}A_{k}\delta(D-D_{c,k})\mu{H}z^2
\label{eq:source}
\end{equation}
where $A_{k}$ and $D_{c,k}$ are adjustable parameters representing the strength and position of the source, respectively, and will be determined by fitting the measured shift data. The numerical factor is arbitrary. Combining equations (\ref{eq:source}), (\ref{eq:ker}), and (\ref{eq:01}) leads to
\begin{equation}
\Delta\nu_{nlm} = \frac{R}{M}\sum^l_{k=0}A_{k}Q_{nl}(D_{c,k})\kappa_{k,lm}
\label{eq:dnu}
\end{equation}
where
\begin{equation}
Q_{nl}(D_{c,k}) = 1.5 \times 10^{-11}\frac{q_{j}(D_{c,k})}{\widetilde{I}_{nl}\nu_{nl}},
\label{eq:Qnl}
\end{equation}

\begin{equation}
\kappa_{nlm} = \int\int|Y^m_{l}|^2P_{2k}(\mu)d\mu{d\phi} = \mathcal{P}^{l}_{2k}(m)Z^l_{k},
\label{eq:09}
\end{equation}
and
\begin{equation}
Z_{k,l} = (-1)^k\frac{(2k-1)!!(2l+1)!!(l-1)!}{k!(2l+2k+1)!!(l-k)!}.
\label{eq:10}
\end{equation}
In these equations, $R$ and $M$ are expressed in solar units, frequencies are expressed in $\mu$Hz, and $\mathcal{P}^{l}_{2k}(m) = lP_{2k}(m/l)$ are orthogonal polynomials of order $2k$ (see Schou et al. \cite{Schou94}).

\subsection{Frequency shifts of $l = 0$ and $l=1$ modes}
\label{sect:radial}
Given the values of $A_{k}$ and $D_{c,k}$, we can evaluate the change in the mode frequencies within individual multiples of low-degree modes using Eq.(\ref{eq:dnu}). However, for radial modes ($l = 0$), both $k$ and $m$ has single zero values, and thus only $A_{0}$ and $D_{c,0}$ are needed. Simple calculation shows $\kappa_{k,lm} = 1$  for $k = l = m = 0$. Then Eq.(\ref{eq:dnu}) becomes
\begin{equation}
\Delta\nu_{n0} = \frac{R}{M}A_{0}Q_{n0}(D_{c,0}).
\label{eq:shift0}
\end{equation}
The values of $A_{0}$ and $D_{c,0}$ can be determined by fitting the measured frequency shifts (see Table \ref{Tab:shifts}) for radial modes. Using least square fitting technique we have
\begin{equation}
A_{0}(D_{c,0}) = \frac{M}{R}\frac{\Sigma_{n}Q_{n0}(D_{c,0})\Delta\nu_{n}^{obs}/(\sigma_{n0}^{obs})^2}{\Sigma_{n}Q_{n0}^2(D_{c,0})/(\sigma_{n0}^{obs})^2}
\label{eq:A0}
\end{equation}
where $\Delta\nu^{obs}_{n0}$ are the measured $l = 0$ shifts and $\sigma^{obs}_{n0}$ are the measured uncertainties. For any given $D_{c,0}$, we can calculate $A_0$ through Eq.(\ref{eq:A0}) and further $\Delta\nu_{n0}$ through Eq.(\ref{eq:shift0}). By brute-force searching from the star interior to the surface, we obtain the best guess of $D_{c,0}$ that minimizes $\chi^2$ (see Fig.\ref{fig:chi2}):
\begin{equation}
\chi^2_{0} = \sum_{n}\bigg(\frac{\Delta\nu_{n0}-\Delta\nu^{obs}_{n0}}{\sigma^{obs}_{n0}}\bigg)^2
 =\sum_{n}\bigg(\frac{\frac{R}{M}A_{0}(D_{c,0})Q_{n0}(D_{c,0})-\Delta\nu^{obs}_{n}}{\sigma^{obs}_{n}}\bigg)^2.
\label{eq:chi2}
\end{equation}
It should be noted that equations (\ref{eq:A0}) and (\ref{eq:chi2}) are different from equations (8) and (9) in Metcalfe et al. (\cite{Metcalfe07}). The later are problematic, because they always lead to too small $A_{0}$ and too large discrepancy between calculated and measured shifts.

If we have measurements of the individual mode frequencies within multiplets, we could use Eq.(\ref{eq:dnu}) directly to calculate $A_k$ and $D_{c,k}$ for \textit{k} up to \textit{l}. However, such measurements are difficult, and we only get the mean frequency shifts (averaged over multiplet components) for $l = 1$ modes as shown in Table \ref{Tab:shifts}. In order to calculate the mean frequency shifts, the following formula is proposed in Dziembowski (\cite{Dziembowski07}):
\begin{equation}\label{eq:shiftl}
  \Delta\nu_{nl} = \frac{2l+1}{2}\frac{R}{M}\sum_{k=0}^{l}\biggl[\sum_{m=-l}^{m=l}|Y_l^m(\theta_0,0)|^2\kappa_{k,lm}\biggr]A_kQ_{nl}(D_{c,k})
\end{equation}

where $\theta_0$ represents the inclination of the rotation axis to the line of sight. For $l = 1$ modes, there are two perturbation sources to contribute to the frequency shifts, corresponding to $k = 0$ and $k = 1$ (see Eq.(\ref{eq:source})), respectively. The above equation becomes
\begin{equation}\label{eq:shift1}
\begin{split}
 \Delta\nu_{n1} &= \frac{3}{2}\frac{R}{M}\sum_{k=0}^{1}\biggl[\sum_{m=-1}^{m=1}|Y_1^m(\theta_0,0)|^2\kappa_{k,1m}\biggr]A_kQ_{n1}(D_{c,k})\\
 &= \frac{9}{8\pi}\frac{R}{M}A_0Q_{n1}(D_{c,0})+\frac{9}{40\pi}\frac{R}{M}(2\cos^2\theta_0-\sin^2\theta_0)A_1Q_{n1}(D_{c,1}).
\end{split}
\end{equation}

Thus, four parameters, $A_0$, $A_1$, $D_{c,0}$ and $D_{c,1}$ in Eq.(\ref{eq:shift1}) need to be determined. $A_0$ and $D_{c,0}$ can be determined through Eq.(\ref{eq:A0}) and (\ref{eq:chi2}) based on the frequency shift of radial modes, while $A_1$ and $D_{c,1}$ are calculated based on the frequency shift of $l = 1$ modes by minimizing the following $\chi^2$ criterion:
\begin{equation}
\chi^2_{1} = \sum_{n}\bigg(\frac{\Delta\nu_{n1}-\Delta\nu^{obs}_{n1}}{\sigma^{obs}_{n1}}\bigg)^2,
\label{eq:chi21}
\end{equation}
with steps analogous to those of calculating $A_0$ and $D_{c,0}$.

\begin{table}
\begin{center}
\caption[]{Evolutionary models (Liu et al. \cite{Liu13}) for HD~49933.}\label{Tab:model}
 \begin{tabular}{ccccccc}
  \hline\noalign{\smallskip}
  \hline\noalign{\smallskip}
     models & M           & $(Z/X)_{s}$  & $T_{eff}$  & $L/L_{\odot}$ &  R/R$_{\odot}$ & age \\
            &(M$_{\odot}$)&              &     (K)    &               &  ($R_{\odot}$) &  (Gyr)\\
     \hline\noalign{\smallskip}
     1 & 1.26 & 0.0061  & 6603 & 3.532 & 1.439 & 1.536 \\
     2 & 1.26 & 0.0073  & 6626 & 3.597 & 1.441 & 1.639 \\
     3 & 1.28 & 0.0097  & 6602 & 3.606 & 1.453 & 1.595 \\
     4 & 1.30 & 0.0092  & 6546 & 3.475 & 1.454 & 1.277 \\
     5 & 1.30 & 0.0106  & 6564 & 3.548 & 1.459 & 1.410 \\
     \hline\noalign{\smallskip}
     6 & 1.26 & 0.0059  & 6608 & 3.540 & 1.452 & 1.580 \\
     7 & 1.26 & 0.0071  & 6632 & 3.614 & 1.442 & 1.696 \\
     8 & 1.28 & 0.0069  & 6557 & 3.467 & 1.445 & 1.402 \\
     9 & 1.28 & 0.0082  & 6583 & 3.540 & 1.448 & 1.521 \\
     10 & 1.28 & 0.0094  & 6608 & 3.614 & 1.452 & 1.646 \\
     11 & 1.28 & 0.0135  & 6537 & 3.491 & 1.458 & 1.837 \\
     12 & 1.29 & 0.0132  & 6577 & 3.595 & 1.462 & 1.763 \\
     13 & 1.30 & 0.0104  & 6568 & 3.483 & 1.459 & 1.454 \\
     \hline\noalign{\smallskip}
     14 & 1.28 & 0.0133  & 6541 & 3.499 & 1.458 & 1.867 \\
     15 & 1.29 & 0.0125  & 6569 & 3.565 & 1.460 & 1.731 \\
     16 & 1.28 & 0.0131  & 6545 & 3.508 & 1.458 & 1.902 \\
     17 & 1.29 & 0.0123  & 6573 & 3.573 & 1.459 & 1.758 \\
  \hline\noalign{\smallskip}\hline
\end{tabular}
\end{center}
\end{table}

\begin{table}
\begin{center}
\caption[]{Computational results of the evolutionary models in Table \ref{Tab:model}.}\label{Tab:results}
 \begin{tabular}{cccccccc}
  \hline\noalign{\smallskip}
  \hline\noalign{\smallskip}
     models & $A_{0}$ & $D_{c,0}$ & $\Delta{i}_{MgII}$$^\ast$  & $\chi_{0}^2$ & $A_1$ & D$_{c,1}$ & $\chi_{1}^2$\\
            &         &        Mm &                            &              &       & Mm        &             \\
     \hline\noalign{\smallskip}
   1 & 13.57 & 0.527 & 0.617 & 2.734 & -3814 & 3.842 & 7.675\\
   2 & 12.93 & 0.516 & 0.588 & 2.699 & -3749 & 3.806 & 7.764\\
   3 & 14.23 & 0.561 & 0.647 & 2.700 & -4149 & 3.868 & 7.810\\
   4 & 15.16 & 0.554 & 0.689 & 2.748 & -4421 & 3.910 & 7.685\\
   5 & 16.55 & 0.620 & 0.752 & 2.735 & -4720 & 3.985 & 7.749\\
   \hline\noalign{\smallskip}
   6 & 13.49 & 0.528 & 0.613 & 2.742 & -3737 & 3.818 & 7.658\\
   7 & 13.20 & 0.534 & 0.600 & 2.702 & -3689 & 3.801 & 7.719\\
   8 & 14.46 & 0.539 & 0.657 & 2.761 & -4084 & 3.872 & 7.638\\
   9 & 14.14 & 0.543 & 0.643 & 2.723 & -4105 & 3.874 & 7.719\\
   10 & 14.18 & 0.564 & 0.645 & 2.707 & -4069 & 3.857 & 7.787\\
   11 & 12.76 & 0.484 & 0.580 & 2.644 & -4133 & 3.745 & 7.933\\
   12 & 14.23 & 0.557 & 0.647 & 2.643 & -4288 & 3.825 & 7.898\\
   13 & 15.14 & 0.577 & 0.688 & 2.736 & -4369 & 3.897 & 7.745\\
   \hline\noalign{\smallskip}
   14 & 13.49 & 0.513 & 0.613 & 2.641 & -4179 & 3.758 & 7.921\\
   15 & 12.91 & 0.504 & 0.587 & 2.672 & -4085 & 3.785 & 7.870\\
   16 & 14.63 & 0.558 & 0.665 & 2.639 & -4333 & 3.809 & 7.886\\
   17 & 13.69 & 0.534 & 0.622 & 2.672 & -4221 & 3.832 & 7.869\\
   \hline\noalign{\smallskip}\hline
   \end{tabular}
   \end{center}
   \tablenotes{\ast}{0.86\textwidth}{$\Delta{i}_{MgII} =A_0 /22$.}
   \end{table}

\section{Numerical results}
\subsection{Radial modes}
\label{sect:Result}
In order to calculate the frequency shifts of HD~49933 using Eq.(\ref{eq:shift0}), a proper stellar model is needed. To reproduce the observed characteristics of HD~49933, we computed in Liu et al.(\cite{Liu13}) a grid of evolutionary tracks with the Yale Rotation Evolution Code (Pinsonneault et al. \cite{Pinsonneault89}; Guenther et al. \cite{Guenther92}; Yang $\&$ Bi 2007a). The initial parameter range of masses and heavy metal abundances are $1.08 - 1.34M_{\odot}$ and $0.006 - 0.030$, separately. Theoretical analysis has been carried out for the star. A total of fifty-four best-fitting models were identified out of hundreds of evolutionary tracks in Liu et al.(\cite{Liu13}), among which parameters of 17 models are listed in Table \ref{Tab:model}. These 17 models can not only reproduce, like other 37 models, the measured temperature, luminosity, and large frequency separation of the star, but also well fit the variation pattern of the small frequency separations in terms of frequencies.

\begin{figure}
  \centering
  \includegraphics{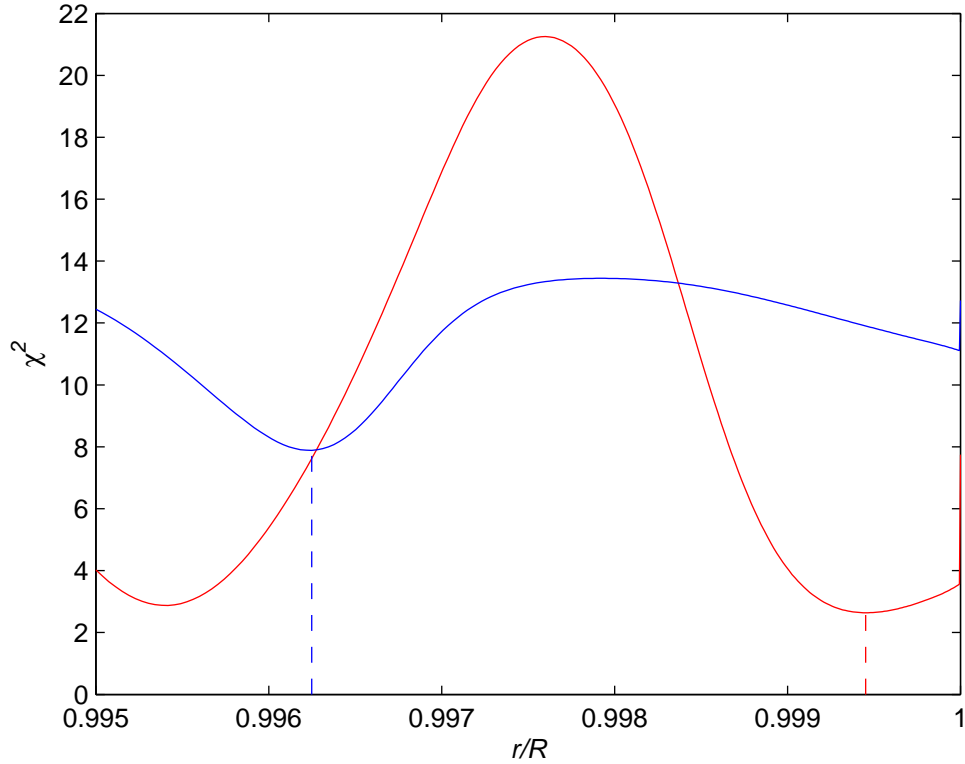}
  \caption{$\chi_0^2$ (red) and $\chi_1^2$ (blue) vs relative radius $r/R$ for model 16. The vertical lines mark the positions $D_{c,0}$ (red) and $D_{c,1}$ (blue) of the source.}
  \label{fig:chi2}
\end{figure}

 The computational results of all 17 models are summarized in Table \ref{Tab:results}. Here, we take model 16 of table \ref{Tab:model} as a representative to our analysis, since it has the smallest $\chi^2_0$ compared to other models.
   Among the 12 measured frequency shifts of radial modes we rule out the last (and largest) shift and only fit the other 11 ones, because this shift is probably an outlier. Values of $\chi_0^2(r)$, defined by Eq.(\ref{eq:chi2}), depend on the stellar radius. The variation of $\chi_0^2(r)$ with radius in the vicinity of the star surface is shown in Fig.\ref{fig:chi2}. Note that in this figure, the $\chi_0^2$ curve has two minima. But the position of the source is unlikely determined by the left minimum, since the right one is globally minimum and the value of $Q_{n0}$
   at the left minimum is far lower than at the right minimum. The best guess of $D_{c,0}$, determined by the right $\chi_0^2$ minimum, is $0.558$ Mm. The corresponding $A_{0}$, obtained by Eq.(\ref{eq:A0}), is $14.63$. The value of $A_{0}$ is much bigger than that of the Sun and $\beta$Hyi obtained by Metcalfe et al. (\cite{Metcalfe07}) (see table \ref{Tab:sunb}). Considering that the measured frequency shifts of HD~49933 are quite large compared to those of the Sun and $\beta$Hyi, this result is not surprising. Because the value of $A_{0}$ is directly related to the variation of magnetic field near the stellar convective zone, we confirm that the magnetic field of HD~49933 varies more drastically through its cycle period than the Sun and $\beta$Hyi.

\begin{figure}
  \centering
  \includegraphics{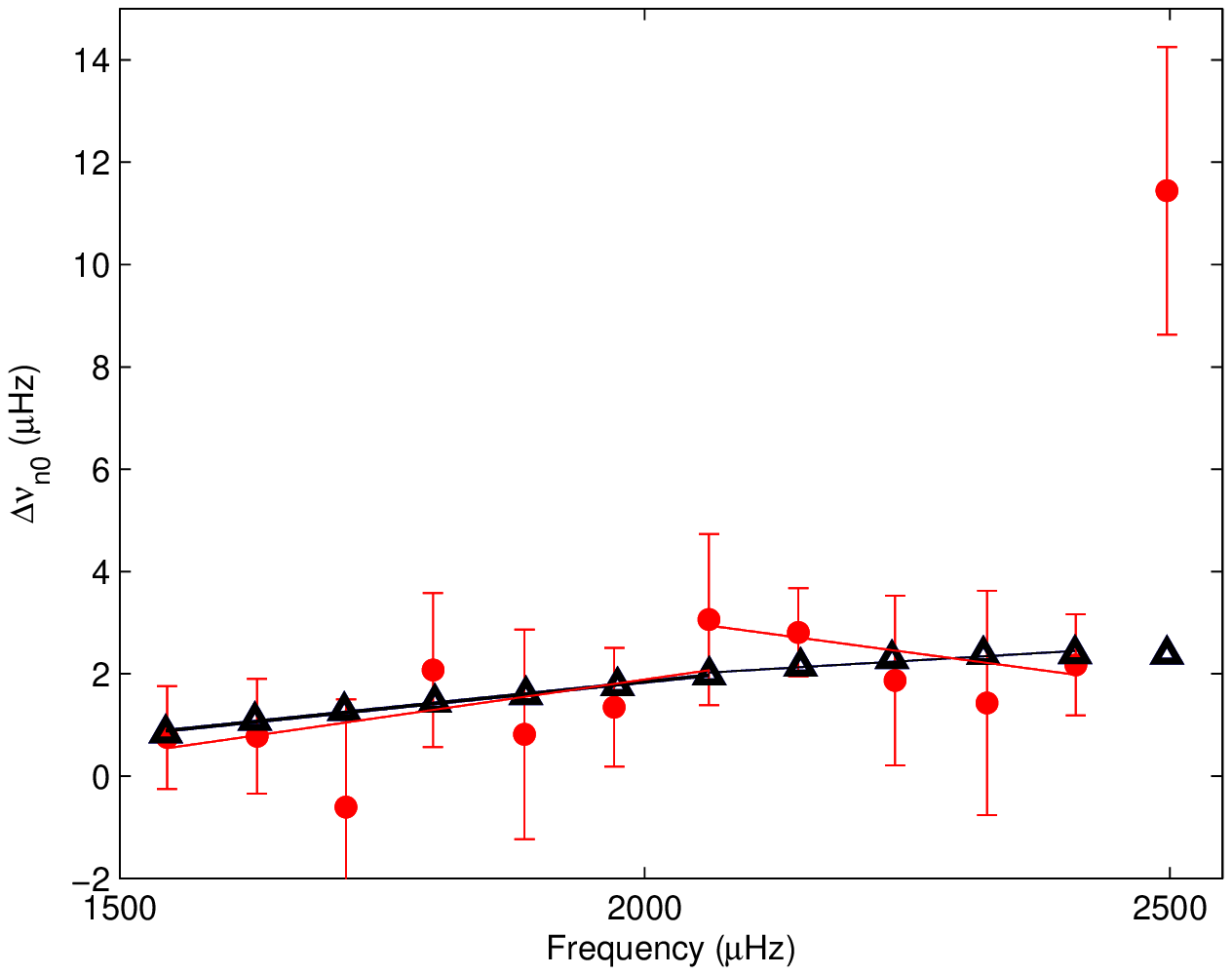}
  \caption{Observed frequency shifts(red dots) for $l = 0$ mode of HD~49933 and theoretical results computed from equation (\ref{eq:dnu}) (black triangles). The red lines correspond to weighted linear fits to the observed shifts, while the black lines correspond to linear fits to the calculated shifts.}
  \label{fig:shift0}
\end{figure}

The frequency shifts calculated from Eq.(\ref{eq:shift0}) for the radial modes of HD~49933, and the observed frequency shifts are shown in Fig.\ref{fig:shift0}. The corresponding least square linear fits, for which the points from the 1st to the 7th and the points from the 7th to the 11th are fitted separately, are also drawn in the figure. A rising trend with increasing frequencies for the calculated shifts is evident, which is consistent with the results of previous work (Goldrich et al. \cite{Goldreich91}; Dziembowski \& Goode \cite{Dziembowski04}; Metcalfe et al. \cite{Metcalfe07}). The measured shifts also grow with frequencies up to 2100 $\mu$Hz (Salabert et al. \cite{Salabert11}), while the four shifts in the range of 2100 - 2400 $\mu$Hz seem to drop subsequently, which cannot reproduced by the present model. However, due to the poor quality of observational data, this dropping pattern is questionable. Generally speaking, the calculations are in good agreement with the measurements. The value of $\chi_0^2$ is 2.639.

The last measured frequency shift, whose value is 11.44 $\mu$Hz, is exceedingly larger than other measured frequency shifts. We cannot get a good fitting with it. Such a two-modal pattern, with different behaviors in low and high frequencies, was also found on frequency shifts of the Sun, for which the fractional frequency shifts rise rapidly at low frequencies and precipitously decline at high frequencies (above $\nu \approx$ 4 mHz). It points to different locations of the sources of the frequency shifts. Since Goldreich et al. (\cite{Goldreich91}) ascribed the sudden decline of the solar frequency shifts at high frequencies to the rise of the solar chromospheric temperature, we naturally reckon that it may be the decrease of the stellar chromospheric temperature that leads to the abrupt rise of the frequency shift of HD~49933. The precipitous nature of the rise results from a chromospheric resonance that occurs at $\nu \approx 2500 \mu$Hz.

Unlike $\beta$Hyi and the Sun, which have been studied in Metcalfe et al. (\cite{Metcalfe07}), the measured MgII index data for HD~49933 are currently not available. Metcalfe et al. (\cite{Metcalfe07}) has got the relationship $A_{0} = 22\Delta{i}_{MgII}$ for the Sun. According to the assumption that the ratio between $A_{0}$ and $\Delta{i}_{MgII}$ is invariant among different solar-like stars (Metcalfe et al. \cite{Metcalfe07}), we can predict that the change of MgII index between minimum and maximum of HD~49933 cycle period should be about 0.665, a number remained to be validated by future observations.

\subsection{Nonradial modes}
  Benomar et al. (\cite{Benomar09}) has measured the inclination angle of HD~49933: $\theta_0 = 17^{\circ\,+7}_{\;\,\:-9}$. According to Eq.(\ref{eq:shift1}), only $A_1$ depends on the value of $\theta_0$, while $\theta_0$ have no impact on $D_{c,1}$ and the resulting $\chi^2_1$. So we randomly set $\theta_0=17^{\circ}$ and calculate $A_1$ and $D_{c,1}$ by fitting the measured frequency shifts of $l = 1$ modes with the $\chi^2$ criterion, i.e. Eq.(\ref{eq:chi21}). In this calculation, we neglect the last observed frequency shifts.

\begin{figure}
  \centering
  \includegraphics{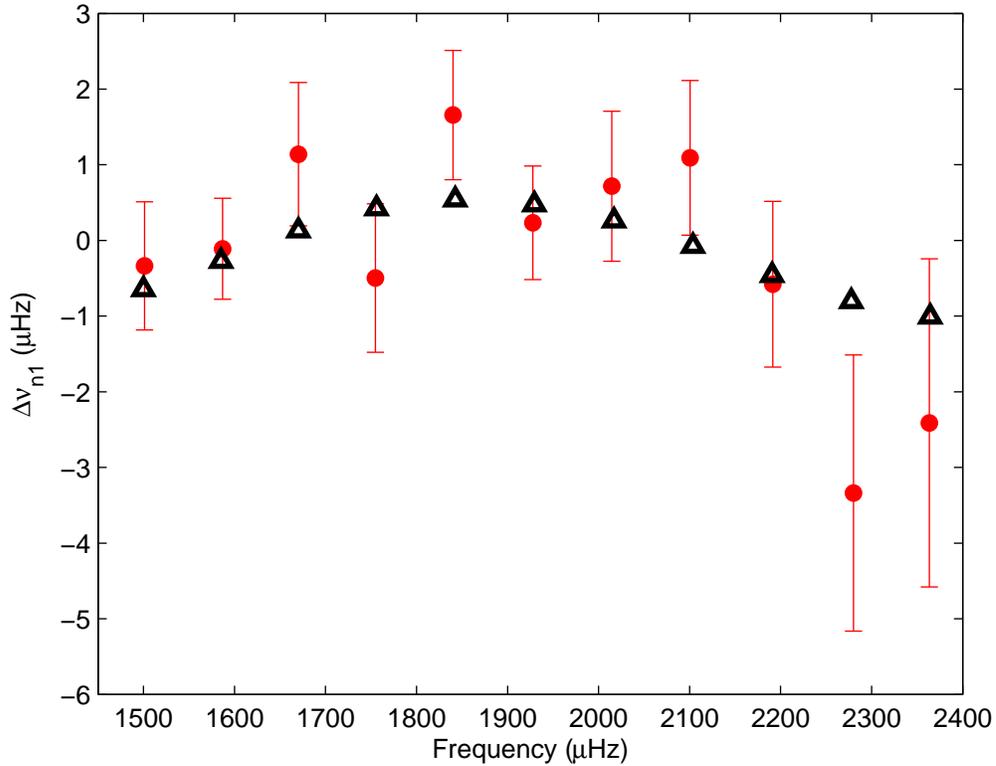}
  \caption{The computed frequency shifts (blue points) and the measured shifts (red points) for $l = 1$ p-modes of HD~49933.}
  \label{fig:shift1}
\end{figure}

The $\chi_1^2(r)$ curve in the vicinity of the star surface is plotted in Fig.\ref{fig:chi2}. Fig.\ref{fig:shift1} illustrates the calculated frequency shifts for the $l = 1$ modes, together with the observed data. We can see that though the agreement between theoretical results and measured shifts for $l = 1$ modes (with $\chi_1^2 = 7.886$) is not as good as that of the radial modes, the general trend of shifts with frequency is reproduced. The best guess of $D_{c,1}$ is 3.809 Mm (see Fig.\ref{fig:chi2}), which means the position of $k = 1$ source is deeper below the stellar surface than the $k = 0$ source. The calculated $A_1$ is negative, with value of -4333. This means the $k = 1$ source may present the variation of turbulent velocity in the outer convective zone. It is commonly accepted that magnetic field impedes convection, and so an increase in magnetic field will reduce turbulent velocity. Dziembowski (\cite{Dziembowski04}) showed that the decrease in the turbulent pressure causes frequency increasing. Therefore, if we assume the variation of turbulent pressure in the stellar convective zone is a type of source, the corresponding $A_1$ should be negative.

\section{Discussion and Conclusions}
\label{sect:conclusion}
In this work, we well reproduce the observed frequency shifts for the radial ($l = 0$) and nonradial ($l = 1$) oscillation modes of a solar-like star HD~49933. Our results show that magnetic active of HD~49933 may be more active than the Sun and $\beta$Hyi, and we predict the change of MgII activity index $\Delta{i_{MgII}}$ between the minimum and maximum of the stellar activity cycle for HD~49933 should be much larger than that observed in the Sun and $\beta$Hyi. Moreover, the position of the source that contributes to both $l = 0$ and $l = 1$ modes is limited in the range $0.48 - 0.62$Mm below the stellar surface.

It is commonly assumed that magnetic fields impede convection, that is, decrease the convective velocity. Our calculation on the frequency shifts of $l = 1$ modes indicates that the decrease of turbulent velocity induced by the increasing magnetic field in the rise phase of HD~49933 activity period may contribute significantly to the $l = 1$ frequency shifts. Based on mixing-length theory, a decrease in the convective velocity is associated with a temperature decrease in the convective zone (Dziembowski $\&$ Goode \cite{Dziembowski05}), which can be reflected by lower stellar effective temperature. Since perturbations of effective temperature and chromospheric temperature are both related to the magnetic field activity, the relationship of the variations of the two types of temperatures with stellar activity cycle is also an issue deserving of research.

\begin{acknowledgements}
We would like to thank D. Salabert for cordially providing us with the measured frequency shifts of HD~49933, and W. A. Dziembowski for his kind guidance for us to understand the method they developed. This work is supported by grants 10933002, 11273007 and 11273012 from the National Natural Science Foundation of China, and the Fundamental Research Funds for the Central Universities.
\end{acknowledgements}

\label{lastpage}

\end{document}